# Highly-Linear Magnet-Free Microelectromechanical Circulators

Yao Yu, *Student Member, IEEE,* Giuseppe Michetti, Ahmed Kord, *Student Member, IEEE,* Michele Pirro, Dimitrios L. Sounas, *Senior Member, IEEE,* Zhicheng Xiao, *Student Member, IEEE,* Cristian Cassella, *Member, IEEE,* Andrea Alù, *Fellow, IEEE,* and Matteo Rinaldi, *Senior Member, IEEE.*

*Abstract*—This paper reports the first demonstration of a magnet-free, high performance microelectromechanical system (MEMS) based circulator. An innovative circuit based on the commutation of MEMS resonators with high quality (Q) factor using RF switches is designed and implemented. Thanks to the high Q factor, a much smaller modulation frequency can be achieved compared to the previous demonstrations, reducing the power consumption and enabling the use of high power-handling switches. Furthermore, the MEMS resonators greatly reduce the required inductance value, guaranteeing much smaller form factor compared to the previous LC demonstrations. The demonstrated circulator shows broad BW (15 dB-IX BW=34.7 MHz for an operational frequency around 2.5 GHz), low IL (4 dB), high IX (30 dB), high linearity (P1dB=28 dBm; IIP3=40 dBm) and at the same time low power consumption, addressing several of the current limitations hindering the full development of magnet-free circulators.

## I. INTRODUCTION

CURRENT wireless communication systems are half-duplex, transmitting and receiving at different time slots or using different carrier frequencies to avoid self-interference between the transmitter (Tx) and the receiver (Rx) modules. Driven by the need of enhancing the spectral efficiency in an overly crowded wireless spectrum, full-duplex operation has been attracting more and more attention over the past few years [1-6]. In full-duplex mode, the wireless terminals transmit and receive signals at the same time over the same frequency band. Circulators, which are three-port non-reciprocal components, are crucial to enable full-duplex operation, due to their ability of prohibiting self-interference between the Tx and Rx nodes without sacrificing insertion loss, since they allow the signal to propagate only in one direction (i.e., from Tx to antenna (ANT) and from ANT to Rx).

Conventional circulators break time-reversal symmetry, and consequently reciprocity, by applying a strong magnetic bias to ferrite cavities [7-9]. However, magnetic-biased circulators are bulky and incompatible with conventional complementary metal-oxide-semiconductor (CMOS) technologies. Active devices, such as transistors, are intrinsically non-reciprocal and hence have been employed to eliminate the requirement on magnetic bias [10-12]. However, these active circulators suffer from poor noise and linearity performances. Recently, a new class of magnet-free circulators based on linear-periodically-time-variant (LPTV) circuits has been proposed [13-37]. In LPTV-based circulators, periodic spatiotemporal modulation is applied to the system to break reciprocity. Based on the components being modulated, LPTV-based circulators can be classified into three categories: transmission line (TL) circulators, lumped element (LC) circulators and MEMS-based circulators. Refs. [13-15] demonstrated TL-based circulators by separating signals being excited from different ports into either different paths using CMOS switches, or different frequencies using distributed modulated varactors. Even though strong non-reciprocity has been achieved with broad BW and low IL, these demonstrations all require large modulation frequencies. This requirement in turn poses three challenges to the system: (i) large power consumption, (ii) leakage of RF power into the modulation paths, and (iii) limited linearity due to the use of either highly-scaled CMOS switches (to improve the switching speed) or solid-state varactors, both of which are limited in linearity. Even though in [13] the linearity from Tx to ANT path was improved by suppressing the voltage swing on the CMOS switches in charge of the modulation, this operation leads to an asymmetrical circulator response and increased sensitivity to impedance mismatch at RF ports, therefore requiring off-chip impedance tuners. Refs. [16]-[18] demonstrated LC circulators based on angular momentum biasing. In these cases, the resonant frequencies of three connected resonant tanks are periodically modulated in a rotating fashion, to break the time-

Manuscript received April 10, 2019. This work was supported by the Defense Advanced Research Projects Agency (DARPA) SPAR program under contract no. HR0011-17-2-0002 and the Air Force Office of Scientific Research.

Y. Yu, G. Michetti, M. Pirro, C. Cassella and M. Rinaldi are with the SMART center, Northeastern University, Boston, MA 02115 USA. (corresponding author: M Rinaldi. Phone: 617-373-2751; Fax: 617-373-8970; E-mail: rinaldi@ece.neu.edu). A. Kord and Z. Xiao are with the Department of Electrical and Computer Engineering, University of Texas at Austin, Austin, TX78712, USA. D. L. Sounas is with the Department of Electrical and Computer Engineering, Wayne State University, Detroit, MI 48202, USA. A. Alù is with the Department of Electrical and Computer Engineering, University of Texas at Austin, TX78712, USA, and also with the Advanced Science Research Center, City University of New York, New York, NY 10031, USA. He is currently the Chief Technology Officer of Silicon Audio RF Circulator. The terms of this arrangement have been reviewed and approved by The University of Texas at Austin and the City University of New York in accordance with its policy on objectivity in research.

reversal symmetry. Low-loss non-reciprocal bands have been achieved with broad BW. Nevertheless limited by the low Q factor of the LC systems, the modulation frequencies required to achieve significant non-reciprocity are at least 10% of RF frequencies. The use of inductors also limits the form factor and integrability of the system. Furthermore, the use of solid-state varactors again limits the systems' linearity and modulation network complexity. Compared to TL or LC components, MEMS resonators have much larger Q factor and better integrability, thus having the potential to greatly reduce the modulation frequency, improve the circulator's insertion loss (IL), isolation (IX) and form factor. In this context, some attempts have been done to implement MEMS circulators [19-22]. However, these demonstrations have all been characterized by severe limitations including high IL, limited IX, narrow bandwidth (BW) and/or intermodulation distortion.

In order to address the aforementioned challenges, we introduce in this paper the first demonstration of a magnet-free circulator using high-Q AlN film bulk acoustic resonators (FBARs) centered at 2.5 GHz, which are spatiotemporally modulated to break time-reversal symmetry. In addition to low IL and high IX, the high Q factor of the system also guarantees much lower modulation frequency requirements (1.6% of RF frequency) compared to previous demonstration, which reduces power consumption and enables the use of high power handling switches. A small modulation frequency generates closely spaced harmonics, which risk to induce significant signal distortion. We address this issue by considering a differential geometry, which has been shown to largely suppress intermodulation mixing [17]. Instead of using conventional varactors, we modulate the center frequencies of the resonant tanks by commuting between two different frequencies using RF switches, therefore achieving better linearity (IIP3=40dBm; P1dB=28 dBm) compared to previous LC demonstrations based on varactors. In fact, thanks to the high Q system and innovative modulation scheme, the achieved linearity is among the highest for all the magnet-free circulators. Finally, in order to improve the BW, the effective electromechanical coupling coefficient ($kt^2$) is increased by connecting inductors in parallel to FBARs to resonate out the static capacitance ($C_0$). Using this principle, we achieve a seven times wider BW. Compared to previous demonstrations based on LC systems, the required inductance values are also greatly reduced (4 nH in this work while hundreds of nH for previous demonstrations), therefore guaranteeing a much smaller form factor and lower cost.

## II. Design

### A. Angular momentum biasing

The reciprocity of a symmetrical three-port resonant network can be broken by applying an angular momentum bias to the system [16]-[19]. As shown in Fig. 1, three resonant tanks with center frequency $f_0$ are connected together, and the center frequencies are modulated periodically with proper phase delay $\varphi_n$, modulation amplitude (i.e. center frequency shift) $a_m$ and angular modulation frequency $\omega_m$. The center frequencies of the three resonant tanks are given by

$$\omega(t) = \omega_0 + a_m \times \cos(\omega_m t + \varphi_n) \qquad (1)$$

where $\omega_0$ is the static center frequency, $\varphi_n$ is the phase of the modulation of the $n$-th tank and $\varphi_1$, $\varphi_2$, $\varphi_3$ are equal to $0^0$, $120^0$ and $240^0$, respectively.

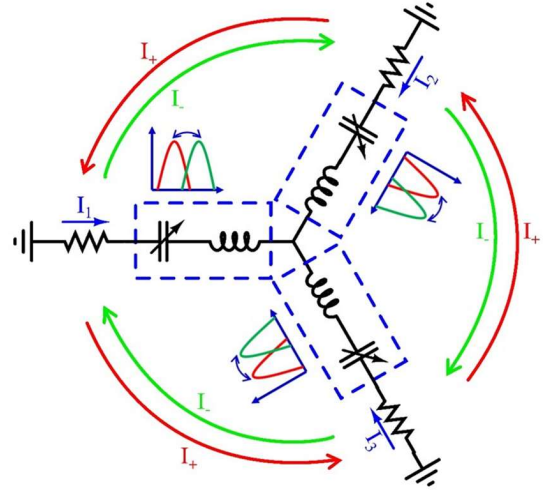

Fig. 1 Schematic of angular momentum biasing through modulation. Three LC tanks are connected to a common node, and the center frequencies of these three tanks are periodically modulated in a rotating fashion to break the degeneracy of the two counter-rotating current mode ($I_-$ and $I_+$) to achieve non-

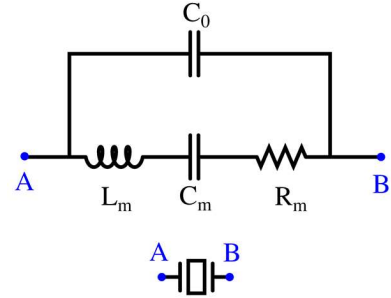

Fig. 2 Equivalent circuit model of an FBAR.

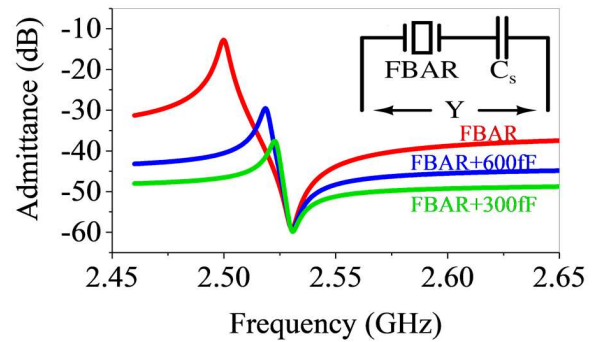

Fig. 3 Circuit simulation of the admittance of an FBAR connected in series to different values of capacitors. Inset: schematic of an FBAR connect in series to a capacitor.

The currents flowing through the three resonant tanks can be seen as a superposition of two counter-rotating modes, $I_-$ and $I_+$

[16-18], as shown in Fig. 1. Without modulation, these two counter-rotating modes are degenerate, therefore if a signal is excited from one port, transmission to the other two ports is equal and the network is reciprocal. However, when modulation is applied to the system with a phase shift specified by (1), a preferred sense of rotation is applied to the rotating modes, since one of them rotates in the same direction as the modulation while the other one rotates in the opposite direction. Therefore, degeneracy is lifted and, by choosing proper modulation amplitude and frequency, the two modes destructively interfere at one port and constructively interfere at the other, thus achieving the operation of a circulator.

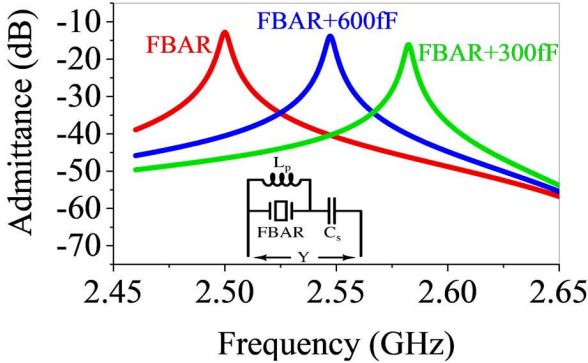

Fig. 4 Circuit simulation of the admittance of an FBAR connected in parallel to an inductor to resonate out $C_0$ and then connect in series to different values of capacitors. Inset: schematic of an FBAR connected in parallel to an inductor $L_p$ and then connected in series to different values of capacitors.

### B. FBAR resonant frequency modulation

The equivalent circuit model of an FBAR is shown in Fig. 2, where $L_m$, $C_m$, $R_m$ and $C_0$ are motional inductance, motional capacitance, motional resistance and static capacitance, respectively. The values of $L_m$, $C_m$ and $R_m$ can be calculated as

$$R_m = \frac{\pi^2}{8} \frac{1}{\omega_0 C_0 kt^2 Q} \qquad (2)$$

$$C_m = \frac{\pi^2}{8} C_0 kt^2 \qquad (3)$$

$$L_m = \frac{\pi^2}{8} \frac{1}{\omega_0^2 C_0 kt^2} \qquad (4)$$

where $\omega_0$ is the center frequency, $kt^2$ is the electromechanical coupling coefficient and Q is the quality factor.

The center frequency of the FBAR is determined by the resonance between $L_m$ and $C_m$, and can be shifted by connecting the FBAR to a series capacitor $C_s$. As shown in [19], as long as $C_m \ll C_0$, the amount of frequency shift can be expressed by

$$\frac{\omega_s' - \omega_s}{\omega_s} = \frac{\Delta \omega_s}{\omega_s} \cong \frac{C_m}{2(C_0 + C_s)} \qquad (5)$$

where $\omega_s$ is the unshifted center frequency and $\omega_s'$ is the shifted center frequency. Furthermore, the introduction of the series capacitor will decrease the admittance magnitude of the FBAR at resonance by a factor $\chi$ [19], i.e.,

$$|Y'_{FBAR}(\omega_s')| = \chi \cdot |Y_{FBAR}(\omega_s)| \qquad (6)$$

Assuming that $1 + \frac{C_s}{C_0} \gg kt^2$ and $(1 + \frac{C_s}{C_0})^2 \ll (kt^2 Q)^2$, the admittance amplitude reduction factor $\chi$ can be expressed as

$$\chi \approx \left(\frac{C_s}{C_s + C_0}\right)^2 \qquad (7)$$

The above analysis highlights two challenges regarding the frequency modulation of FBARs compared to the case of simple LC resonators: first, according to (5), the modulation amplitude is limited by $C_m/C_0$, i.e., the $kt^2$ of FBARs, which is the reason why all the previous MEMS-resonator based circulator demonstrations showed narrow BW. Second, the frequency shift will cause a reduction in admittance at resonance, which is also determined by the value of $C_0$, thus causing higher IL of the circulator. Fig. 3 shows the simulated results of frequency shift when an FBAR is connected to a series capacitor. The FBAR used in simulation is assumed to have a $C_0$ of 1 pF $kt^2$ of 3%, Q of 600 and $\omega_0$ of $2\pi \times 2.5$ GHz. Consistent with the above analysis, the modulation amplitude is limited by the anti-resonance peak determined by the $kt^2$ of the FBAR, and the admittance at peak drops with the decrease of $C_s$.

The two challenges can be simultaneously addressed by decreasing the value of $C_0$, or equivalently, increasing the $kt^2$ of the FBAR resonators. However, the $kt^2$ of the FBAR is determined by the piezoelectric coefficient of AlN and has a theoretical upper bound of ~7% for FBARs. Therefore, in this paper, in order to address the two challenges, inductors are connected in parallel to the FBARs. The value of the parallel inductors is chosen such that they can resonate with $C_0$ at the center frequencies of the FBARs, thus we will have a larger effective $kt^2$ around the center frequency. Therefore, when the center frequencies are shifted, the admittance at resonance will reduce by a much smaller ratio.

Fig. 4 shows the simulated results of frequency shift when an FBAR is connected in parallel to an inductor. As explained before, the value of $L_p$ is chosen to be $L_p = \frac{1}{\omega_0^2 C_0}$. As expected, with the same values of series capacitance $C_s$, the modulation amplitudes are significantly increased, and the peak admittance reduction is much smaller.

### C. Circulator design

The circulator circuit combines two single-ended (SE) branches connected in a differential configuration with a modulation phase difference of $180^0$ (Fig. 5a). This configuration was proven to have the ability to cancel intermodulation products [17], therefore improving IL and IX. This is particularly important in this implementation, given the close proximity of these mixing products to the signal frequency. For each SE branch, three FBAR resonators are connected in a differential configuration. In order to simplify the printed circuit board (PCB) implementation, the three parallel inductors ($L_p$) are connected in a delta configuration, instead of wye. Using wye-to-delta transformation, this is equivalent to connecting them in wye configuration with three times smaller inductance (Fig. 5b). Instead of modulating the center frequencies by using varactors, three capacitors ($C_s$) are connected in series to the three FBARs respectively, and RF switches are connected in parallel to series capacitors and are modulated by square waves. Therefore, each of the resonant tank is commutated between two center frequencies. By synchronizing the modulation for each of the SE branch to have an increase of phase of $120^0$ towards either clockwise (CW) or counterclockwise (CCW) direction, a preferred rotation direction is formed and therefore, degeneracy of the two counter rotating modes is lifted and reciprocity is broken. Thanks to the ultra-low modulating frequency enabled by the use of high Q FBARs, switches with high power handling can be used. Compared to the conventional modulation mechanism using varactors, the use of highly-linear RF switches guarantees high linearity and much simplified modulation network.

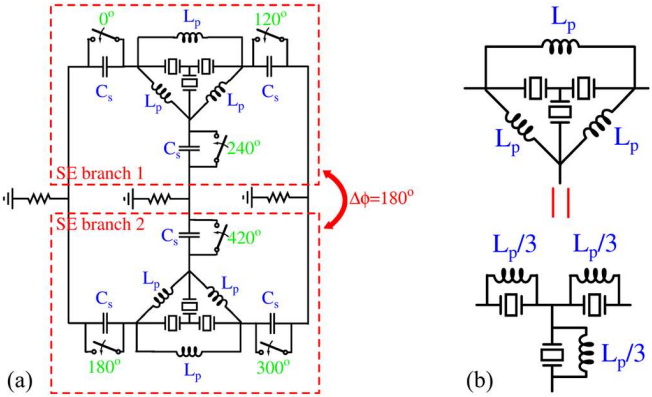

Fig. 5 (a) Circuit schematic of the proposed magnet-free circulator. The circuit contains two single-ended branch, with a modulation phase difference of $180^0$. For each of the single-ended branch, three FBARs are connected in a wye configuration, and the three parallel inductors are connected in a delta configuration, for easier PCB implementation. The FBARs are connected in series to switched capacitors to modulate the center frequency. (b) The equivalence of the connection of parallel inductors in wye and delta configuration. A three times larger inductance is needed for the delta connection.

The value of series capacitors is chosen by circuit simulation, shown in Fig. 6. The circuit is simulated using the harmonic balance simulator in Keysight ADS. The IL and BW (defined by the BW at 15 dB of IX) versus series capacitance are plotted. For each of the series capacitance, the modulation frequency is chosen such that the maximum IX is 25 dB. When the capacitance is too small, the admittance reduction at resonance causes a high IL. When the capacitance is too large, the modulation amplitude is not enough to provide significant IX and low IL at the same time. The optimal value of series capacitance lies between 200 fF to 300 fF. On the other hand, BW increases with smaller series capacitance and therefore larger modulation amplitude. Therefore, the value of series capacitance in this paper is chosen to be 200 fF, in order to achieve low IL and high BW at the same time. Using the optimal value of series capacitance, circuit simulation shows an IL of 3.6 dB, IX of 25 dB and 15 dB-BW of 28 MHz (Fig. 7).

It is worth mentioning that the quality factor Q of the parallel inductors does not degrade the performance of the circuit too much. The simulated relationship between the circulator IL and Q of $L_p$ is plotted in Fig. 8. Compared to Q of 500, using inductors with Q of 125 (commercially available) will only increase the IL by 0.1 dB.

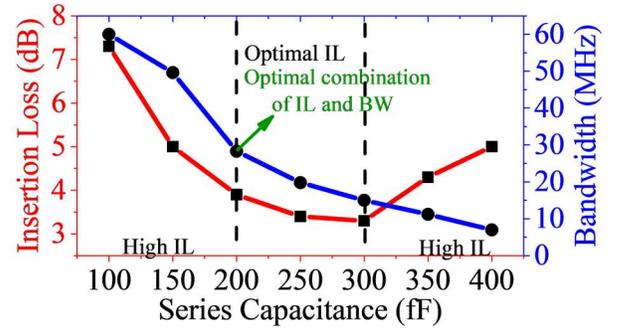

Fig. 6 Simulated results of IL and BW with different series capacitance values. For each of the capacitance value, modulation frequency is set such that the IX is 25 dB.

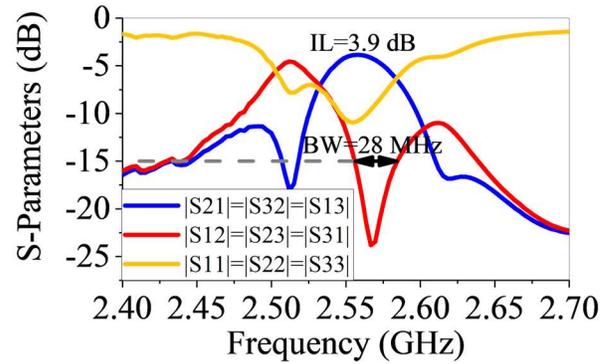

Fig. 7 Circuit simulation of S-parameters with the optimum series capacitance (200 fF). Simulated results show a IL of 3.9 dB, IX of 25 dB and 15 dB-IX BW of 28MHz.

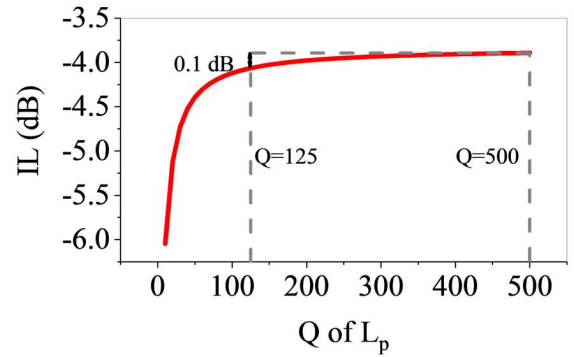

Fig. 8 The relationship between IL and Q of the parallel inductors. Compared to a Q of 500, using a Q of 125 (commercially available) will only degrade the IL by 0.1 dB.

### D. PCB design

A PCB was designed and fabricated in order to implement the described circulator (Fig. 9a). FBARs used in this paper (Fig. 9b) (Broadcom engineering sample) are monolithically integrated, showing a center frequency of 2.5 GHz, $kt^2$ of ~3% and Q of ~600 (Fig. 9c) and are wire-bonded to the PCB.

Parallel inductors are Coilcraft 0603HP series with inductance of 11.92 nH and Q of 125 at 2.5 GHz. RF switches are MACOM MASWSS0179 single-pole, double-throw (SPDT) switches with IL of ~1 dB at 2.5 GHz. Series capacitors are Murata GJW series with capacitance of 200 fF.

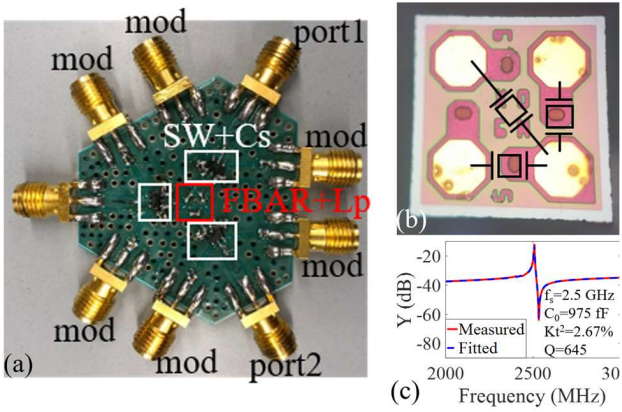

Fig. 9 (a) Designed PCB for the circulator. (b) FBAR chip (Broadcom engineering sample) and schematic of connection. Three FBARs are monolithically integrated on one chip. (c) The MBVD fitting of the FBARs. Result shows a $C_0$ of 975 fF, $kt^2$ of 2.67% and Q of 645.

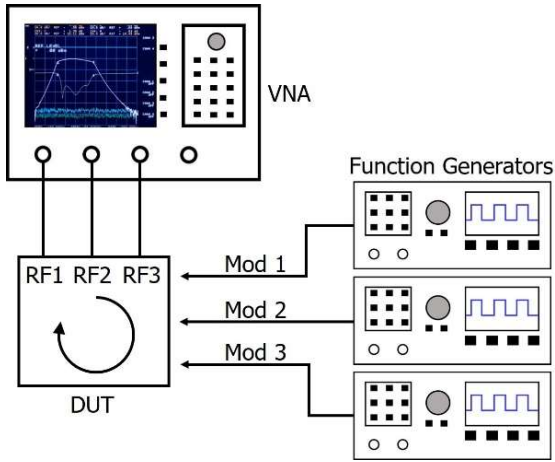

Fig. 10 Measurement set up. Three function generators are synchronized together to provide the modulation signal. The S-parameters of the device are measured using a 4-port VNA.

## III. MEASUREMENTS

### A. S-parameters

The measurement of the S-parameters is shown in Fig. 10. In order to test the PCB, three two-channel function generators are synchronized to provide the square wave modulation signals. The modulation frequency is set to be 40 MHz. The S-parameters are measured using a 4-port vector network analyzer (VNA).

The measurement of S-parameters shows IL of 4.5, 4.7 and 5.3 dB; IX of 24.5, 21.1 and 20.3 dB and RL of 23.2, 18.7 and 13.2 dB, respectively (Fig. 11). The slight difference is attributed to the random values fluctuations of components, parameters mismatch between different FBARs and imperfect phase synchronization of modulation signals. Furthermore, if port 1, 2, and 3 are connected to Rx, Tx and ANT, the IL of signal transmission from ANT to Rx and from Tx to ANT and the IX between Tx and Rx, which are the most important metrics in full-duplex operation, can be optimized by slightly tuning the modulation phase. Fig. 12 (a) reports the optimized S-parameters, showing an IL of 4.0 dB, IX of 30 dB, and 15 dB-IX BW of 34.7 MHz (~1.4% of RF frequency).

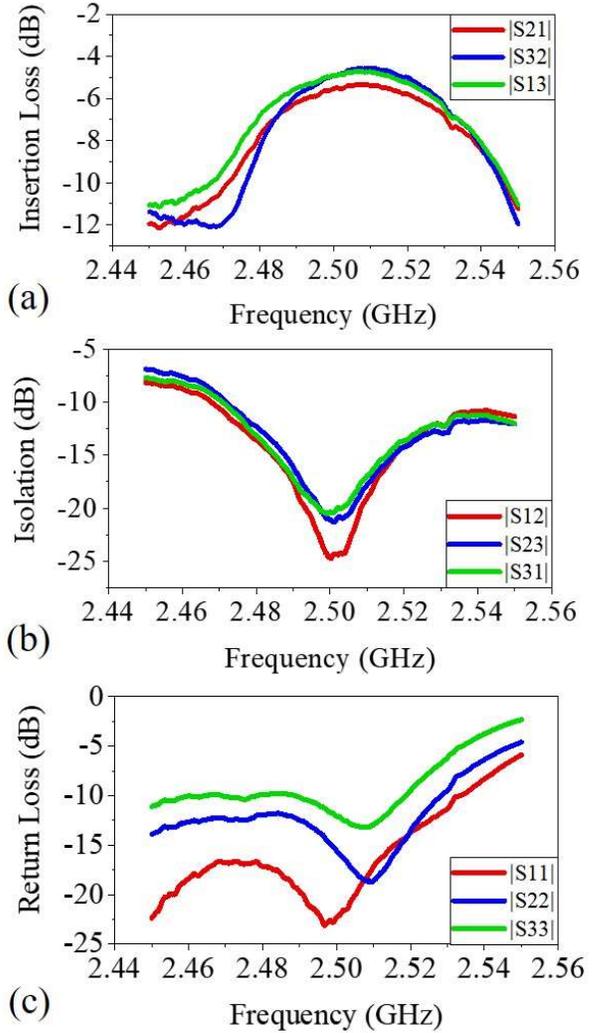

Fig. 11. Measured S-parameters of the circuit when the modulation phases are set to be exactly $120^0$ between each other. (a)Measured IL. (b) Measured IX. (c) Measured RL.

The S-parameters without parallel inductors are also measured (Fig. 12b). Due to a much smaller modulation amplitude, a lower modulation frequency is used (7 MHz). As expected, the BW is almost 7 times narrower, and the IL is also higher, due to the limited modulation amplitude and large admittance reduction factor determined by $C_0$. Note that in this case a switch with a lower IL (0.5 dB at 2.5 GHz) but faster switching speed is used (MASWSS0166), due to a less rigorous requirement on the switching speed.

### B. Output spectrum

The output spectrum is measured using a spectrum analyzer (Fig. 13). Single tone signal is excited from port 1, and the output spectrum is measured at port 2, while port 3 is terminated by 50 ohm impedance. The measured output spectrum shows more than 30 dBc of first-order intermodulation products suppression, thanks to the use of differential configuration. The reason of imperfect intermodulation suppression is attributed to slight asymmetry of PCB designs and components values fluctuation between differential paths, and small modulation phase mismatch.

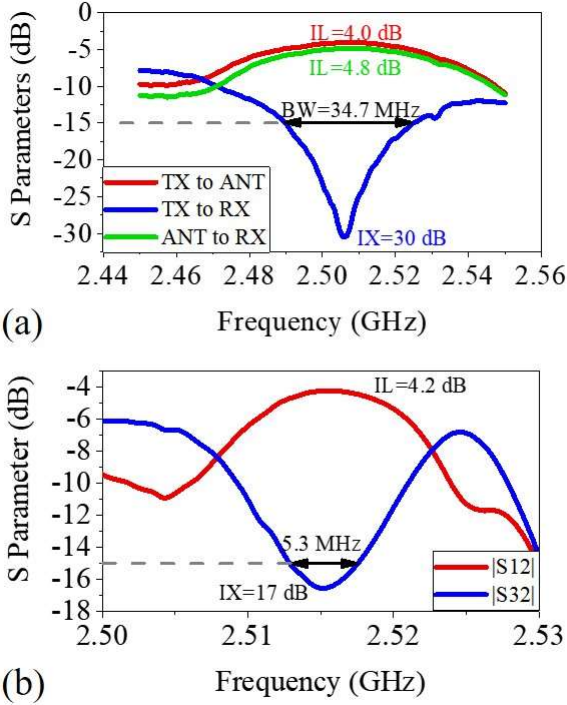

Fig. 12. (a) Measured IL from TX to ANT and from ANT to RX and measured IX from TX to RX. The modulation phases are slightly tuned to optimize these three parameters. (b) Measured S-parameters of the circulator without parallel inductors. A 7 times narrower BW is observed.

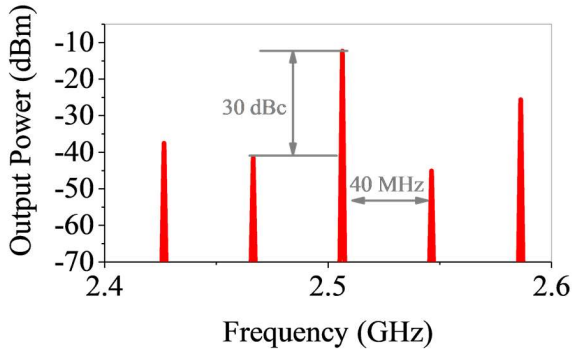

Fig. 13. Measured output spectrum

### C. Linearity and power handling

The linearity is evaluated by measuring the 1 dB compression point (P1dB) and input-referred third-order harmonic intersect point (IIP3). The circuit shows excellent linearity. Measured P1dB is 28 dBm and IIP3 is 40 dBm (Fig. 14). The measured linearity is among the highest for all magnet-free circulators demonstrated to date, thanks to the use of RF switches with high power handling, enabled by the ultra-low modulation frequency.

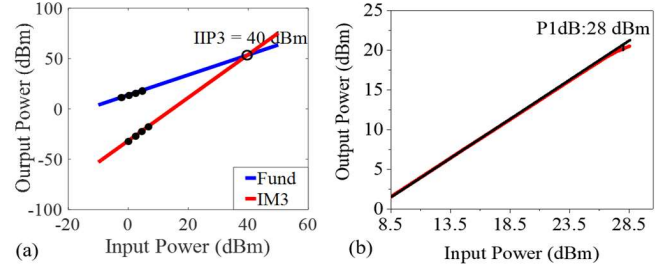

Fig. 14. Measured Linearity (a) IIP3 (b) P1dB.

### IV. CONCLUSION

Table I summarizes the performance metrics reported in this work compared to other relevant works on magnet-free circulators based on LPTV circuits. Among these demonstrations, Refs. [13] and [25] are based on the spatiotemporal modulation of TLs, Refs. [17] and [18] are implemented through angular momentum biasing using LCs, while Refs. [19], [20], [22], [26] and this work are based on spatiotemporal modulation of MEMS devices. In this work, for the first time without sacrificing other performance metrics including IL, BW and linearity, ultra-low modulation frequency (1.6%) has been achieved, thanks to the use of high Q FBARs. Besides a much smaller power consumption, this low modulation frequency also enables the use of RF switches with high power handling, therefore leading to a power handling performance (P1dB of 28 dBm and IIP3 of 40 dBm) that is among the highest for all magnet-free circulator demonstrations. The achieved IL (4.0 dB) is much lower than all the other MEMS based circulators, and is also among the lowest compared to all previous work at a relatively high center frequency (2.5 GHz), again due to the use of high Q FBARs centered at 2.5 GHz. Compared to previous demonstrations based on varactors [17], [18], the use of RF switches significantly simplifies the modulation network and improves the linearity. Furthermore, by using parallel inductors to increase the effective $kt^2$ of the FBARs, a broad BW (1.4%) is achieved, overcoming the narrowband issue from previous MEMS demonstrations. In summary, this paper shows the first demonstration of a 2.5 GHz highly-linear and broadband FBAR-based magnet-free circulator that shows low IL, high IX and low power consumption at the same time. The demonstrated response shows the potential towards high-performance RF non-reciprocal component with extremely small form factor that can be integrated in modern communication systems to achieve full-duplex operation.

### ACKNOWLEDGMENTS

The authors thank Dr. Rich Ruby from Broadcom Limited for providing FBARs engineering samples.

TABLE I
COMPARISON TO OTHER LPTV CIRCULATORS

| | Technology | Center freq. | Mod. Freq. [a] | BW [b] | IX | IL | P1dB | IIP3 |
|---|---|---|---|---|---|---|---|---|
| [13] | TL | 25 GHz | 33% | 18.4% | 18.3 dB | 3.3/3.2 dB | 21.5/21 dBm | N/A |
| [25] | TL | DC-3GHz [c] | 83% [d] | 93.3% [e] | 20 dB | 4.3 dB | N/A | N/A |
| [18] | LC | 1000 MHz | 19% | 2.4% | 20 dB | 3.3 dB | 29 dBm | 34 dBm |
| [17] | LC | 1000 MHz | 10% | 2.3% | 20 dB | 0.8 dB [f] | 29 dBm | 32 dBm |
| [22] | MEMS | 155 MHz | 0.6% | 5.8% | 20 dB | 6.6 dB | N/A | 30 dBm |
| [20] | MEMS | 2500 MHz | 0.1% | 0.02% | 20 dB | 11 dB | N/A | N/A |
| [19] | MEMS | 146 MHz | 0.1% | 0.2% | 15 dB | 8 dB | -8 dBm | N/A |
| [26] | MEMS | 1165 MHz | 0.1% | 0.3% | 15 dB | 12 dB | N/A | N/A |
| **This work** | **MEMS** | **2500 MHz** | **1.6%** | **1.4%** | **15 dB** | **4.0 dB** | **28 dBm** | **40 dBm** |

[a-b] Defined by the ratio with center frequency.
[b] Defined by the IX value.
[c] Results are broadband measured from DC to 3 GHz.
[d-e] Assuming center frequency is 1.5 GHz.
[f] Baluns are de-embedded

# REFERENCES

[1] J. Choi I, M. Jain, K. Srinivasan, P. Levis, and S. Katti, "Achieving single channel, full duplex wireless communication," *In Proc. 16th Annu. Int. Conf. Mobile Computing and Networking*, 2010, pp. 1-12.

[2] M. Jain, J. Choi I, T. Kim, D. Bharadia, S. Seth, K. Srinivasan, P. Levis, S. Katti and P. Sinha, "Practical, real-time, full duplex wireless," *In Proc. 17th Annu. Int. Conf. Mobile Computing and Networking*, 2011, pp. 301-312.

[3] D. Bharadia, E. McMilin and S. Katti, "Full duplex radios," *ACM SIGCOMM Comput. Commun. Rev.*, vol. 43, no. 4, pp. 375-386, Sept. 2013.

[4] D. W. Bliss, P. A. Parker and A. R. Margetts, "Simultaneous transmission and reception for improved wireless network performance," *In Proc. IEEE/SP 14th Workshop Stat. Signal Process.*, pp. 478-482, 2007.

[5] M. Duarte and A. Sabharwal, "Full-duplex wireless communications using off-the-shelf radios: Feasibility and first results," in *Proc. Conf. Rec. 44th ASILOMAR Conf. Signals, Syst. Comput.*, pp. 1558-1562, 2010.

[6] M. Duarte, C. Dick and A. Sabharwal, "Experiment-driven characterization of full-duplex wireless systems," *IEEE Trans. Wireless Commu.*, Vol. 11, no.12, 4296-4307, 2012.

[7] D. M. Pozar, "Microwave engineering", John Wiley & Sons, 2009.

[8] C. E. Fay and R. L. Comstock, "Operation of the ferrite junction circulator," *IEEE Trans. Microw. Theory and Techn.*, vol.13, no. 1, pp. 15-27, 1965.

[9] H. Bosma, "On stripline Y-circulation at UHF," *IEEE trans. Microw. Theory and Techn.*, vol. 12, pp. 61-72, 1964.

[10] S. Tanaka, N. Shimomura and K. Ohtake, "Active circulators: The realization of circulators using transistors," *In Proc. IEEE*, vol. 53, no.3, pp. 260-267, 1965.

[11] Y. Ayasli, "Field effect transistor circulators," *IEEE Trans. On Magnetics*, vol. 25, no. 5, pp. 3243-3247, 1989.

[12] S. A. Ayati, D. Mandal, B. Bakkaloglu and S. Kiaei, "Adaptive integrated CMOS circulator," *Radio Frequency Integrated Circuits Symposium*, pp. 146-149, 2016.

[13] T. Dinc, M. Tymchenko, A. Nagulu, D. Sounas, A. Alu and H. Krishnaswamy, "Synchronized conductivity modulation to realize broadband lossless magnetic-free non-reciprocity," *Nat. Commun.*, vol. 8, no. 1, 795, 2017.

[14] S. Qin, Q. M. Xu and Y. E. Wang, "Nonreciprocal components with distributedly modulated capacitors," *IEEE trans. Microw. Theory and Techn.*, vol. 62, no.10, pp. 2260-2272, 2014.

[15] M. M. Biedka, R. Zhu, Q. M. Xu and Y. E. Wang, "Ultra-wide band non-reciprocity through sequentially-switched delay lines," *Scientific reports*, vol. 7, 40014, 2017.

[16] N. A. Estep, D. L. Sounas, J. Soric and A. Alù, "Magnetic-free non-reciprocity and isolation based on parametrically modulated coupled-resonator loops," *Nature Physics*, vol.10, no.12, 923, 2014.

[17] A. Kord, D. L. Sounas and A. Alù, "Pseudo-Linear Time-Invariant Magnetless Circulators Based on Differential Spatiotemporal Modulation of Resonant Junctions," *IEEE trans. Microw. Theory and Techn.*, vol. 66, no. 6, pp. 2731-2745, Jun. 2018.

[18] A. Kord, D. L. Sounas and A. Alù, "Magnet-less circulators based on spatiotemporal modulation of bandstop filters in a delta topology," *IEEE trans. Microw. Theory and Techn.*, vol. 66, no. 2, 911-926, 2018.

[19] Y. Yu, G. Michetti, A. Kord, D. Sounas, F. V. Pop, P. Kulik, M. Pirro, Z. Qian, A. Alu and M. Rinaldi, "Magnetic-free radio frequency circulator based on spatiotemporal commutation of MEMS resonators," *In Proc. IEEE Micro Electro Mechanical Systems (MEMS)*, pp. 154-157, 2018.

[20] M. M. Torunbalci, T. J. Odelberg, S. Sridaran, R. C. Ruby and S. A. Bhave, "An FBAR Circulator," *IEEE Microwave and Wireless Components Letters*, vol. 28, no. 5, 395-397, 2018.

[21] C. Xu, E. Calayir and G. Piazza, "Magnetic-free electrical circulator based on AlN MEMS filters and CMOS RF switches," *In Proc. IEEE Micro Electro Mechanical Systems (MEMS)*, pp. 755-758, 2018.

[22] R. Lu, T. Manzaneque, Y. Yang, L. Gao, A. Gao, and S. Gong, "A Radio Frequency Non-reciprocal Network Based on Switched Acoustic Delay Lines," *IEEE trans. Microw. Theory and Techn.*, to be published.

[23] A. Kord, D. L. Sounas, Z. Xiao, and A. Alu, "Broadband Cyclic-Symmetric Magnet-less Circulators and Theoretical Bounds on their Bandwidth," *IEEE trans. Microw. Theory and Techn.*, vol. 66, no. 12, pp.5472-5481.

[24] A. Nagulu, Andrea Alù, and Harish Krishnaswamy, "Fully-Integrated Non-Magnetic 180nm SOI Circulator with> 1W P1dB,>+ 50dBm IIP3 and High Isolation Across 1.85 VSWR," *In 2018 IEEE Radio Frequency Integrated Circuits Symposium (RFIC)*, pp. 104-107, 2018.

[25] M. Biedka, Q. Wu, X. Zou, S. Qin, and Y. E. Wang, "Integrated time-varying electromagnetic devices for ultra-wide band nonreciprocity," *In Proc. 2018 IEEE Radio and Wireless Symp. (RWS)*, pp. 80-83. IEEE, 2018.

[26] C. Xu, and G. Piazza, "Magnet-Less Circulator Using AlN MEMS Filters and CMOS RF Switches," *J. Microelectromech. Syst.*, to be published.

[27] R. Lu, J. Krol, L. Gao, and S. Gong. "A Frequency Independent Framework for Synthesis of Programmable Non-reciprocal Networks," *Scientific reports*, vol. 8, no. 1, 14655, 2018.

[28] A. Kord, D. L. Sounas, and A. Alù, "Achieving Full-Duplex Communication: Magnetless Parametric Circulators for Full-Duplex Communication Systems," *IEEE Microw. Mag.*, vol. 19, no. 1, pp. 84-90, Jan. 2018.

[29] M. Pirro, C. Cassella, G. Michetti, G. Chen, P. Kulik, Y. Yu, and M. Rinaldi. "Novel topology for a non-reciprocal MEMS filter," *In Proc. 2018 Int. Ultrasonics Symp. (IUS)*, pp. 1-3. IEEE, 2018.

[30] G. Michetti, C. Cassella, F. Pop, M. Pirro, A. Kord, D. Sounas, A. Alú, and M. Rinaldi. "A quasi-LTI frequency-selective SAW circulato," *In Proc. 2018 IEEE Int. Ultrasonics Symp. (IUS)*, pp. 206-212. IEEE, 2018.

[31] M. Ghatge, G. Walters, T. Nishida, and R. Tabrizian. "A Non-Reciprocal Filter using Asymmetrically Transduced Micro-Acoustic Resonators," *IEEE Electron Device Lett.*, to be published.

[32] N. Reiskarimian, A. Nagulu, T. Dinc, and H. Krishnaswamy, "Nonreciprocal Electronic Devices: A Hypothesis Turned Into Reality," *IEEE Microwave Magazine*, vol. 20, no. 4, 94-111, 2019.

[33] Y. Yu and M. Rinaldi. "Wi-Fi-band Acoustic Resonant Circulator," *In Proc. IEEE Micro Electro Mechanical Systems (MEMS), 2019.*

[34] P. Lulik, Y. Yu, G. Chen and M. Rinaldi, "Magnetic-free Isolator Based on a Single Differential Lattice Filter, " *In Proc. IEEE Micro Electro Mechanical Systems (MEMS), 2019.*

[35] Y. Yu, F. V. Pop, G.Michetti, P. Kulik, M. Pirro, A. Kord, D. Sounas, A. Alu and M. Rinaldi, "2.5 GHz Highly-Linear Magnetic-Free

Microelectromechanical Resonant Circulator," *In Proc. IEEE Int. Freq. Cont. Symp, 2018.*

[36] D. Sounas, and A. Alù, "Non-Reciprocal Photonics Based on Time Modulation," *Nature Photonics*, Vol. 11, No. 12, pp. 774-783, November 30, 2017.

[37] Y. Yu, and M. Rinaldi, "Chip-scale micro-acoustic radio frequency gyrator," *In. Proc. IEEE Int. Conf. Solid-State Sensors, Actuators and Microsystems (Transducers),* 2019.